\documentclass{IEEEtran}
\usepackage{cite}
\usepackage{amsmath,amssymb,amsfonts}
\usepackage{algorithmic}
\usepackage{graphicx}
\usepackage{textcomp}
\usepackage{graphicx,epstopdf}
\usepackage{stfloats}
\usepackage{csquotes}

\usepackage{subfigure}
\usepackage{amsmath}
\usepackage{tabularx}
\usepackage{footnote}
\usepackage{adjustbox}
\usepackage{amsthm}
\usepackage{setspace}
\usepackage{tikz}
\usetikzlibrary{%
    decorations.pathreplacing,%
    decorations.pathmorphing%
}
\usetikzlibrary{shapes.geometric, arrows}

\tikzstyle{process1} = [rectangle, minimum width=2cm, minimum height=0.8cm, text centered, text width=2.1cm, draw=black,fill=gray!10]
\tikzstyle{process2} = [rectangle, minimum width=1.5cm, minimum height=1cm, text centered, text width=1.5cm, draw=black, fill=gray!10]
\tikzstyle{process3} = [rectangle, minimum width=3cm, minimum height=0.5cm, text centered, text width=3cm, draw=black, fill=gray!10]
\tikzstyle{process} = [circle, minimum width=2.5cm, minimum height=0.5cm, text centered, text width=2.5cm, draw=black, fill=gray!10]
\tikzstyle{decision} = [diamond, minimum width=3cm, minimum height=0.5cm, text centered, draw=black, fill=green!30]
\tikzstyle{arrow} = [thick,->,>=stealth]

\def\BibTeX{{\rm B\kern-.05em{\sc i\kern-.025em b}\kern-.08em
    T\kern-.1667em\lower.7ex\hbox{E}\kern-.125emX}}
\begin{document}
\title{Modelling of Vertical Dipole Above Lossy Dielectric Half-Space: Characteristic Mode Theory}
\author{Sandip Ghosal,  Arijit De, Raed~M.~Shubair, and Ajay Chakrabarty
}
\maketitle

\maketitle

\begin{abstract}
\color{black}This work introduces a theoretical extension of the characteristic mode formulation for analysing the vertical electric dipole lying above a lossy dielectric half-space. As the conventional characteristic formulation fails to maintain the orthogonality of the characteristic field modes over the infinite sphere, an alternate modal formulation is proposed here to maintain the orthogonality for both the current and field modes. The modal results are found to match closely with its method of moment counterparts. Later, the modes of an isolated dipole with no ground plane have been used to predict the role of the lossy ground plane through a theory of the linear combination of the eigenvectors. The proposed formulations have been studied with different heights from the ground plane and are compared with the direct modal solutions to validate its accuracy. It helps to provide a thorough understanding of how the isolated modes interact among each other to constitute the perturbed modes in the presence of the lossy half-space. It can find application to include the lossy earth effect in the study of the lightning fields and the path loss modelling of the antennas over the lossy ground.
 \color{black}
\end{abstract}


\newpage
\section{Introduction}
\label{sec:introduction}
\color{black}
\IEEEPARstart{T}{he} radiation from the vertical dipole lying over the lossy ground seems to constitute the starting point of the electromagnetic fields due to the lightning phenomena which can be analysed as a superposition of multiple vertical dipoles at different heights from the ground plane \cite{zou2010fast}, and \cite{lee2011fast}. Apart from the lightning applications, an array of vertical dipoles is a common use case for beam scanning or spatial modulation. It was found in \cite{qian1990impedance} that the lossy nature of the ground plane has a significant influence on the input impedance of the dipoles, which control the scanning behavior. Alternately, the path loss variation of the dipole also depends on the ground plane, as found in \cite{fei2007comparative} and \cite{de2010characterization}. The initial solution of the electromagnetic fields due to the vertical electric dipole(VED) above the dielectric half-space was proposed in \cite{sommerfeld1926ausbreitung}. However, the improper integral of \cite{sommerfeld1926ausbreitung} has the limitations of oscillating nature and slow convergence of the upper integral limit. Therefore, different approximations were proposed for computing the horizontal field due to the VED in \cite{cooray2002some}. In a similar way, different asymptotic techniques were suggested to speed up the computation process in \cite{zou2010fast} and \cite{lee2011fast}. However, most of such approaches assume that the radiating source is an electrically small vertical dipole. Therefore, an alternate method of moment (MOM) based theory was proposed for the vertical dipole with finite length in \cite{yang1991discrete} and \cite{shubair1993closed} using discrete complex image techniques. It was found that only a few terms in  Prony's method can provide good convergence in \cite{yang1991discrete} and \cite{shubair1993closed}. The discrete complex image technique is found to be suitable for the array of dipoles in \cite{qian1990impedance}.    
The present work also considers a similar integral approach to compute the MoM impedance matrix, but a generalized eigen decomposition technique is adopted instead of the matrix inversion. One main motivation for adopting the characteristic mode analysis (CMA) is that the characteristic modes are independent of the excitation vector. In addition, computation of the modes at a single frequency can provide the information of the modes around the neighbouring frequency region.\\
\color{black}The CMA was initially proposed by Garbacz et al. in \cite{garbacz1971generalized}. Later, the method of moment (MoM) of the electric field integral equation (EFIE) was used to define the current based formulation of characteristic modes in \cite{harrington1971theory} for the perfect electrically conducting (PEC) elements. The CMA of dielectric and magnetic bodies were reported in \cite{harrington1972characteristic}. The CMA has gained profound attention in the recent two decades, specifically for the antenna design \cite{lau2016guest} and \color{black}for analysing the radiation physics of electromagnetic interference (EMI) as found in \cite{cao2017quantifying}, \cite{cao2017emi}, and \cite{yang2018characterizing}.
The CMA for the multilayer structures were reported in \cite{chen2015characteristic} and \cite{qi2018multi} where only lossless cases had been considered. When there is a finite loss in structure, the conventional characteristic relation of \cite{harrington1971theory}, \cite{chen2015characteristic}, and \cite{qi2018multi} can no longer provide a set of orthogonal characteristic fields. It provides the motivation for exploring the characteristic mode analysis for the PEC dipoles in the presence of a lossy dielectric half-space. \color{black}Considering the role of the lossy ground plane in the lightning analysis \cite{zou2010fast}, and \cite{lee2011fast}, or in the placement of antennas in wireless propagation \cite{sghosalthesis}, this work extends the preliminary analysis of the existing work of \cite{ghosal2018antem1} in a more extensive and detailed way. The discrete complex image approach of \cite{shubair1993closed} is followed here for computing the MoM impedance matrix. Later, a new characteristic equation is proposed for the generalized eigen decomposition of the impedance matrix. The modal results are compared with the MoM counterparts. In the next, the role of the lossy ground has been studied using the linear combination theory of eigenvectors of the isolated dipole  \cite{TCCM1}. It helps to provide a clear understanding of the evolution of the isolated modes into the perturbed modes in the presence of the lossy half-space.\\
The work is arranged in the following way. Section II discusses the modal analysis of the PEC dipole above lossy half-space. Section III proposes a theory of linear combination to analyze the effects of the lossy ground on the characteristic current distribution of the dipole. A generalized formulation, as well as a first-order closed-form solution, have been derived for the modal analysis of the vertical dipole above the lossy dielectric half-space. The concluding section summarises the total works and highlights its possible scope of application. The work of this project adds to the previous contributions \cite{khan2016compact,khan2016compact1,khan2017ultra,omar2016uwb,shubair2015novel1,al2006direction,nwalozie2013simple,shubair2005robust,belhoul2003modelling,ibrahim2017compact,shubair2004robust,che2008propagation,el2016design,al2005computationally,al2003investigation,shubair2005performance,al2016millimeter,shubair2005convergence,ibrahim2016reconfigurable,ghosal2018characteristic,khan2018novel,elsalamouny2015novel,khan2016pattern,8458184,7156466,7696430,7347957,8646600,7305294,8706527,7803806,7803808,9082193,alhajri2018accurate,al2005direction,al2003performance} reported in the literature in new applications and evolving technologies.

\newpage
\section{Characteristic Mode Formulation}
\label{sec:guidelines}
\begin{figure}[ht!]
\centering
\includegraphics[height=6cm]{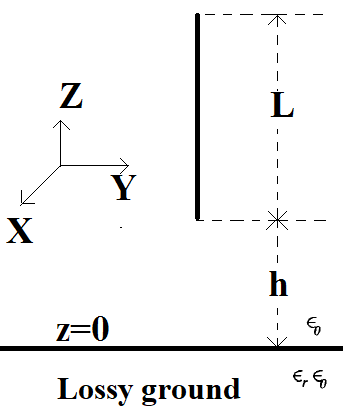}
\caption{PEC dipole above the infinite lossy dielectric ground plane.}
\label{fig3}
\end{figure}

\color{black}
For illustration, let an isolated Z-directed thin PEC dipole be considered with the length of $\text{L}$ in Fig. \ref{fig3}. The height from the lossy dielectric ground plane is $h$. The free-space permittivity is denoted by $\epsilon_0$. For the lossy dielectric half-space, corresponding relative permittivity $\epsilon_r$ will be a complex number. As the tangential electric field on the surface of the PEC dipole is zero, corresponding current density $\text{J}_{LG}$ is related to the excitation field $\text{E}$ through the following Pocklington's formulation \cite{shubair1993closed}
\begin{align}
    \text{E}(z)=\frac{1}{\text{j}\omega\epsilon_0}\int_{h}^{L+h}(k_0^2+\frac{\partial^2}{\partial z^2})G_{LG} (z,z') J_{LG}(z)dz'.
    \label{eqn1}
\end{align}
As the Z-directed dipole is assumed to be thin with respect to the corresponding wavelength ($\lambda$), only the Z-component ($\text{J}_{LG}(z)$) of the current density will hold significant value. Here, $k_o=\frac{2\pi}{\lambda}$ indicates the free-space wavenumber. When the dipole is vertically lying above the lossy dielectric half-space, there appear multiple numbers of complex images in its equivalent representation in \cite{shubair1993closed}. As a consequence, the Green's function $G_{LG} (z,z')$ needs to consider the contribution of $M$ number of complex images in \eqref{eqn1}. Therefore, the Green's function $G_{LG}(z,z')$ will be
\begin{align}
    G_{LG}(z,z')&=\frac{e^{-jk_0 d_o}}{4\pi d_o}-K_{eps}\frac{e^{-jk_0 d_{i}}}{4\pi d_{i}}+\sum_{m=1}^M u_m\frac{e^{-jk_0 d_{cm}}}{4\pi d_{cm}}. 
\label{eqn16}
\end{align}
For an isolated PEC dipole lying in the free-space without the ground plane, corresponding only the first term will be present in \eqref{eqn16} where the distance vector between the source ($x',y',z'$) and the observation ($x, y, z$) locations is denoted by $d_o$ as
\begin{align}
    d_o&=\sqrt{(x-x')^2+(y-y')^2+(z-z')^2}\nonumber\\
    &=\sqrt{\rho_o^2+(z-z')^2}
    \label{eqn3}
\end{align}
As the dipole is Z-directed, its distance vector of the quasi-dynamic image will be \cite{shubair1993closed}
\begin{align}    
d_{i}=\sqrt{(x-x')^2+(y-y')^2+(z+z')^2}\nonumber\\=\sqrt{\rho_o^2+(z+z')^2}
\label{eqn10}
\end{align}
The distance vector of the $m^{th}$ complex image will be
\begin{align}
d_{cm}&=\sqrt{\rho^2+(z+z'-jv_m)^2}, K_{eps}=\frac{1-\epsilon_r}{1+\epsilon_r}.
\label{eqn17}
\end{align}
As $\epsilon_r$ is a complex number for the lossy dielectric ground, the lossy coefficient $K_{eps}$ will also be complex. Alternately, for the infinite PEC ground, the coefficient $K_{eps}$ will be $-1$, and there will be no formation of complex image \cite{shubair1993closed}. The contribution from the complex images of the equivalent current is considered by the weighted sum of the third term in \eqref{eqn16}. Following \cite{shubair1993closed}, Prony's method will be followed here to determine the complex expansion coefficients $u_m$ and $v_m$. Considering the Galerkin's matching technique \cite{harrington1993field}, the MoM equation can be written for the dipole above lossy ground as
\begin{align}
[\text{V}]&=[\text{Z}_{LG}][\text{J}_{LG}].
\label{eqn18}
\end{align}
\subsection{Conventional Formulation}
If conventional CMA of \cite{harrington1971theory} is followed to compute the eigenmodes, corresponding characteristic equation of the impedance matrix $[\text{Z}_{LG}]=[\text{R}_{LG}]+\text{j}[\text{X}_{LG}]$ will be
\begin{align}
[\text{X}_{LG}]\tilde{J}_{LG,n}=\Delta_{LG,n}[\text{R}_{LG}]\tilde{J}_{LG,n}.
\label{eqn19}
\end{align} 
The eigenvectors $\tilde{J}_{LG,n}$ and eigenvalues $\Delta_{LG,n}$ will be real in \eqref{eqn19}. Its orthogonal relationship can be written as
\begin{align}
[\tilde{J}_{LG,m}]^T[\text{X}_{LG}][\tilde{J}_{LG,n}]&=\Delta_{LG,n}[\tilde{J}_{LG,m}]^T[\text{R}_{LG}][\tilde{J}_{LG,n}]\nonumber\\&=\delta_{mn}\Delta_{LG,n}[\tilde{J}_{LG,n}]^T[\text{R}_{LG}][\tilde{J}_{LG,n}],\label{eqn20}
\end{align} 
\begin{align}
\text{where}~~\delta_{mn}=0,~\text{if}~m \neq n,~\text{else}~0.
\label{eqn21}
\end{align} 
Due to the presence of the lossy coefficient $K_{eps}$, the matrix $[\text{R}_{LG}]$ will have both the free-space radiation and lossy dissipation counterparts. The radiation part ($[\text{R}_{o}]$) is due to the first term in \eqref{eqn16}. The other two terms contribute to the lossy part of $[\text{R}_{L}]=[\text{R}_{LG}]-[\text{R}_{o}]$.\\
The introduction of characteristic mode analysis is fundamentally rooted in the far-field scattering matrix where the far-field scattering matrix was decomposed into standard eigen decomposition as $[S]E_n=\lambda^S_n E_n$ in \cite{garbacz1965modal} and \cite{garbacz1971generalized} to obtain a set of characteristic field modes $E_n$ which are orthogonal over the far-field radiation sphere. Later, the MoM implementation of the electric field integral equation on the surface (/volume) of the lossless objects were used in \cite{harrington1971theory} to find out a set of characteristic current modes which can generate a set of characteristic field modes which maintain the orthogonality over the infinite radiation sphere. Such inter-relationship of the current based CMA and the initial field-based CMA can be noted in \cite{harrington1971theory} and \cite{harrington1972characteristic}. So, the succeeding discussion checks whether the conventional characteristic mode formulation of \eqref{eqn19} can maintain such orthogonality properties in the set of characteristic field modes.\\ 
Let the $n^{th}$ eigenvector $\tilde{J}_{LG,n}$ be assumed to produce the the $n^{th}$ characteristic field pattern $\tilde{E}_{LG,n}$ in the far-field. If the complex power balance equation is followed using \cite{harrington1961time} and \cite{harrington1971theory}, it can be written that
\begin{align}
[\tilde{J}_{LG,m}]^T[\text{Z}_{LG}][\tilde{J}_{LG,n}]=\frac{1}{\eta_o}\oint_{S\infty}\tilde{E}_{LG,m}^*.\tilde{E}_{LG,n} dS\nonumber\\+[\tilde{J}_{LG,m}]^T[\text{R}_{L}][\tilde{J}_{LG,n}] +\text{j}\omega\int_{V'}(\mu \tilde{H}_{LG,m}^*.\tilde{H}_{LG,n}\nonumber\\-\epsilon_o\tilde{E}_{LG,m}^*.\tilde{E}_{LG,n}) dV.
\label{eqn22a}
\end{align}

Separating the real part of \eqref{eqn22a}, it can be written as
\begin{align}
\frac{1}{\eta_o}\oint_{S\infty}\tilde{E}_{LG,m}^*.\tilde{E}_{LG,n} dS +[\tilde{J}_{LG,m}]^T[\text{R}_{L}][\tilde{J}_{LG,n}]\nonumber\\
=\delta_{mn}[\tilde{J}_{LG,n}]^T[\text{R}_{LG}][\tilde{J}_{LG,n}].
\label{eqn22b}
\end{align}
Here, the free-space intrinsic impedance is $\eta_o$. The inner product of the characteristic
field modes $\tilde{E}_{LG,m}$ and $\tilde{E}_{LG,n}$ over the far-field radiation sphere ($S\infty$) provides the radiated power. For $m\neq n$, \eqref{eqn22b} can be written using \eqref{eqn20} as
\begin{align}
\frac{1}{\eta_o}\oint_{S\infty}\tilde{E}_{LG,m}^*.\tilde{E}_{LG,n} dS=-[\tilde{J}_{LG,m}]^T[\text{R}_{L}][\tilde{J}_{LG,n}]
,~m\neq n.
\label{eqn22c}
\end{align}
For the lossless case, as $[\text{R}_{L}]=[0]$, the right hand term of \eqref{eqn22c} will be zero, providing the orthogonality to the field modes $\tilde{E}_{LG,m}$ and $\tilde{E}_{LG,n}$. However, in the presence of the lossy dielectric ground, the resistive power loss term $\text{P}_{RL~m,n}=|[\tilde{J}_{LG,m}]^T[\text{R}_{L}][\tilde{J}_{LG,n}]|$ can be non-zero. As a consequence, the
orthogonal relationship of the modal field patterns $\tilde{E}_{LG,n}$ at far-field location will not be maintained as shown in \eqref{eqn22c}. It physically signifies that the characteristic current modes $\tilde{J}_{LG,n}$ fail to diagonalize the far-field scattering matrix. So, the least-square convergence property over infinite sphere is also lost \cite{harrington1971theory}. The situation is very similar to eigenmodes of the lossy dielectric bodies in  \cite{harrington1972characteristic}. 
\subsection{Proposed Formulation}
As noted in \eqref{eqn22c}, the projection of the eigenvectors in the matrix space of $[\text{R}_{L}]$ leads to the loss of orthogonality in the far-field components. To maintain the far-field orthogonal property \cite{harrington1972characteristic}, an alternate formulation is provided in the present work for computing the characteristic modes of the PEC dipole above lossy ground. The modes can be determined using the following characteristic relation,
\begin{align}
 [\text{X}_{LG}][\text{J}_{LG,n}]&=\lambda_{LG,n}[\text{R}_{o}][\text{J}_{LG,n}].
  \label{eqn25}
\end{align}
As both $[\text{X}_{LG}]$ and $[\text{R}_{o}]$ are symmetric and $[\text{R}_{o}]$ is positive definite, the eigenvectors $[\text{J}_{LG,n}]$ and eigenvalues $\lambda_{LG,n}$ will be real. The orthogonal property of the $n^{th}$ characteristic mode can be expressed as
\begin{align}
[\text{J}_{LG,m}]^T[\text{X}_{LG}][\text{J}_{LG,n}]=\delta_{mn}\lambda_{LG,n}[\text{J}_{LG,m}]^T[\text{R}_{o}][\text{J}_{LG,n}].
\label{eqn26}
\end{align}
Similar to the conventional case, the complex power balance equation can be written using \cite{harrington1961time} and \cite{harrington1971theory}
\begin{align}
[\text{J}_{LG,m}]^T[\text{Z}_{LG}][\text{J}_{LG,n}]=\frac{1}{\eta_o}\oint_{S\infty}\tilde{E}_{LG,m}^*.\text{E}_{LG,n} dS\nonumber\\ +[\text{J}_{LG,m}]^T[\text{R}_{L}][\text{J}_{LG,n}] +\text{j}\omega\int_{V'}(\mu \tilde{H}_{LG,m}^*.\text{H}_{LG,n}\nonumber\\-\epsilon_o\text{E}_{LG,m}^*.\text{E}_{LG,n}) dV.
\label{eqn27a}
\end{align}
Separation of the real part of \eqref{eqn27a} will provide
\begin{align}
\frac{1}{\eta_o}\oint_{S\infty}\text{E}_{LG,m}^*.\text{E}_{LG,n} dS +[\text{J}_{LG,m}]^T[\text{R}_{L}][\text{J}_{LG,n}]\nonumber\\
=[\text{J}_{LG,m}]^T[\text{R}_{o}][\text{J}_{LG,n}]+[\text{J}_{LG,m}]^T[\text{R}_{L}][\text{J}_{LG,n}]\nonumber\\
=\delta_{mn}[\text{J}_{LG,n}]^T[\text{R}_{o}][\text{J}_{LG,n}]+[\text{J}_{LG,n}]^T[\text{R}_{L}][\text{J}_{LG,n}].
\label{eqn27b}
\end{align}
For $m \neq n$, it can be seen from \eqref{eqn27b} that
\begin{align}
\frac{1}{\eta_o}\oint_{S\infty}\text{E}_{LG,m}^*.\text{E}_{LG,n} dS 
=\delta_{mn}[\text{J}_{LG,n}]^T[\text{R}_{o}][\text{J}_{LG,n}]=0.
\label{eqn27c}
\end{align}
Thus, the characteristic modes obtained using the proposed formulation of \eqref{eqn27b} provide a set of characteristic field modes $\text{E}_{LG,m}$ and $\text{E}_{LG,n}$ which are orthogonal over the far-field radiation sphere. As a comparison between the conventional CMA formulation of \eqref{eqn19} and the proposed formulation of \eqref{eqn25}, it is worthy of mentioning the following points: 
\begin{itemize}
\item In both the methods of \eqref{eqn19} and \eqref{eqn25}, the dimension of the total eigenspace does not change.
\item The characteristic modes and eigenvalues are real for both the methods of \eqref{eqn19} and \eqref{eqn25}.
\item The conventional method of \eqref{eqn19} fails to provide the far-field orthogonality. Therefore, the total scattered field in the far-field region can not be expressed as a weighted linear combination of the characteristic field modes obtained using the conventional characteristic relation of \eqref{eqn19}.
\item The proposed formulation of \eqref{eqn25} can maintain the orthogonality relation of both the characteristic current modes and the characteristic field modes. Lastly, the formulation is also applicable for the lossless case too when $[\text{X}_{LG}]=[\text{X}_L]+[\text{X}_o]=[\text{X}_o]$ and $[\text{R}_{LG}]=[\text{R}_L]+[\text{R}_o]=[\text{R}_o]$.
\item As the number of characteristic modes is the same \eqref{eqn19} and \eqref{eqn25}; the convergence behaviour will be controlled by the modal weighting coefficients only. The convergence of the total current or field depends on the proper selection of dominant modes, as discussed in more detail in the previous works \cite{sghosalthesis} and \cite{TCCM1}.
\end{itemize}
The total induced current density $\text{J}_{LG}$ can be expressed as a weighted linear combination of the characteristic current modes $\text{J}_{LG,n}$
\begin{align}
\text{J}_{LG}=\sum_{n=1}^N \alpha_{LG,n}\text{J}_{LG,n}.
\label{en1}
\end{align}

\begin{figure}[ht!]%
\centering
\subfigure[]{%
\label{fig1b}%
\centering
\includegraphics[height=2.0in, width=2.2in]{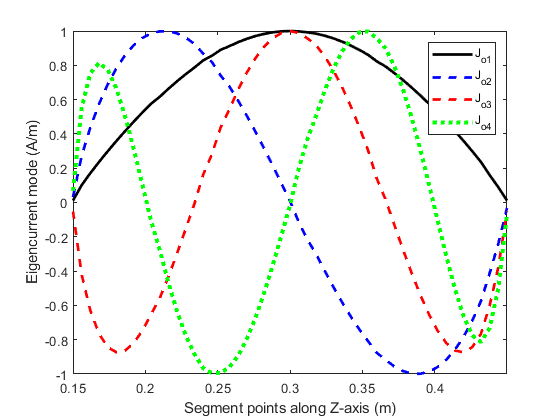}}
\hfill
\subfigure[]{%
\label{fig1c}%
\centering
\includegraphics[height=2.0in, width=2.2in]{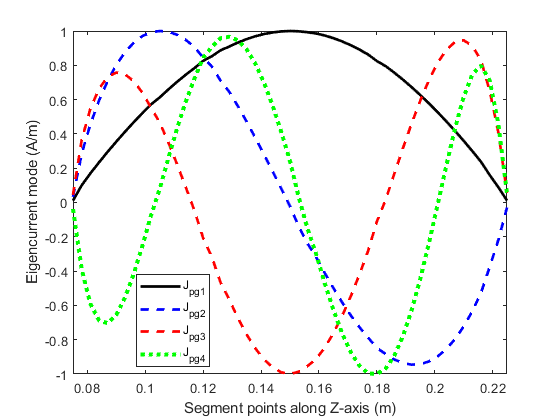}}
\hfill
\subfigure[]{%
\label{fig1d}%
\centering
\includegraphics[height=2.0in, width=2.2in]{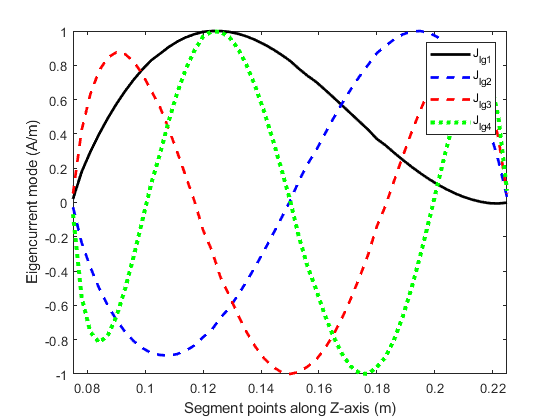}}
\caption{(a) Isolated PEC dipole in the free-space without the ground plane, (b) dipole at $h=\lambda/4$ above the PEC half-space, (c) the PEC dipole above lossy dielectric half-space with $h=\lambda/4$; $\epsilon_r=16-j16$.}
\end{figure}

The column vector $[\alpha_{LG}]$ modal weighting coefficient $\alpha_{LG,n}$ can be found out as
\begin{align}
[\alpha_{LG}]_{N \times N}=\begin{bmatrix}
\alpha_{LG,1}\\
\alpha_{LG,2}\\
\vdots\\
\alpha_{LG,N}
\end{bmatrix}=[\text{P}^{LG}_{mn}]_{N \times N}^{-1}[\text{V}^{LG}_{n}]_{N \times N},
\label{en2}
\end{align}
where
\begin{align}
[\text{P}^{LG}_{mn}]=
\begin{bmatrix}
\text{P}^{XG}_{11}+\text{P}^L_{11} & \text{P}^L_{11} & \hdots &\text{P}^L_{1N}\\
\text{P}^L_{21} & \text{P}^{XG}_{22}+\text{P}^L_{22} & \hdots &\text{P}^L_{2N}\\
\vdots & \vdots & \hdots & \vdots\\
\text{P}^L_{N1} & \text{P}^L_{N2} & \hdots & \text{P}^{XG}_{NN}+\text{P}^L_{NN}
\end{bmatrix}.
\label{en3}
\end{align}
The modal powers $\text{P}^{XG}_{mn}$ and $\text{P}^{L}_{mn}$ are defined as
\begin{align}
\text{P}^{XG}_{mn}=(1+\text{j}\lambda_{LG,m})[\text{J}_{LG,m}]^T[\text{R}_{o}][\text{J}_{LG,n}],\nonumber\\
\text{P}^{L}_{mn}=[\text{J}_{LG,m}]^T[\text{R}_{L}][\text{J}_{LG,n}]~\text{and}~\text{V}^{LG}_{n}=[\text{J}_{LG,n}]^T[\text{V}].
\label{en4}
\end{align}
The next section provides a comparative discussion how the characteristic modes vary for three different cases: Case A-Isolated PEC dipole without the ground plane, Case B- PEC dipole lying above the PEC half-space and Case C-PEC dipole lying above the lossy dielectric half-space.\\
For the sake of comparison, a Z-directed half-wavelength (L= $\lambda/2$) PEC thin dipole is considered without the ground plane. Its first four dominant characteristic current modes computed using \eqref{eqn6} at 1 GHz are shown in Fig. \ref{fig1b}. Here, the free-space wavelength at 1 GHz is denoted by $\lambda$. The modes for the same half-wavelength dipole with an infinite PEC ground plane using \eqref{eqn13} are shown in Fig. \ref{fig1c}. It can be seen from Figs. \ref{fig1b} and \ref{fig1c} that the infinite PEC ground plane maintains the shape of the modal distribution which matches with the study of a previous literature \cite{cabedo2007theory} where PEC plate was taken as an example.

\begin{table}[ht!]
\begin{center}
\caption{Comparison of the eigenvalues for the three cases with $h=\lambda/4$}
\label{Table1}
\begin{tabular}{|c|c|c|c|}
 \hline
Mode index & Case A & Case B & Case C\\
 \hline
 1 & $0.6638$ &$0.6913$ &$2.2617$\\
 \hline
 2 & $-117.1075$ &$-89.9118$ &$-103.7868$\\
 \hline
 3 & $-7.3837E+3$ &$-7.4529E+3$ &$-7.4532E+3$\\
 \hline
 4 & $-7.9505E+5$ &$-7.9504E+5$ &$-7.9488E+5$\\
 \hline
 \end{tabular}
\end{center}
\end{table}

 The infinite PEC ground plane causes a shift in the resonating frequency in \cite{cabedo2007theory}. As the eigenvalue reflects the resonance in characteristic mode analysis \cite{harrington1971theory}, the eigenvalues are found to be changed in Table \ref{Table1}. However, the lossy ground causes a significant change in the modal distribution as shown in Fig.\ref{fig1d} and Table \ref{Table1}.
 
\begin{table}[ht!]
\centering
\small
\caption{Error of the eigenvalues for different cases with $h=\lambda/4$.}
\label{Table2}
\begin{tabular}{|c|c|c|c|}
 \hline
 $\delta \lambda_{n}$ & Case-2  & Case-3 \\
 \hline
 $\delta \lambda_{1}$ & $4.1428$ & $240.7158$\\
 \hline
 $\delta\lambda_{2}$ & $23.2229$ & $11.3748$\\
 \hline
  $\delta\lambda_{3}$  & $0.9373$ & $0.9409$\\
 \hline
 $\delta\lambda_{4}$  & $0.0013$ & $0.0213$\\
 \hline
\end{tabular}
\end{table}

\begin{table}[!ht]
\centering
\small
\caption{Error of the eigenvectors for different cases with $h=\lambda/4$.}
\label{Table3}
\begin{tabular}{|c|c|c|}
 \hline
 $\delta \angle \text{J}_{n}$ & Case-2  & Case-3 \\
 \hline
 $\delta \angle \text{J}_{1}$ & $0.1684$ & $23.0979$\\
 \hline
 $\delta \angle \text{I}_{2}$ & $2.1291$ & $3.3648$\\
 \hline
  $\delta \angle\text{J}_{3}$  & $5.0717$ &  $0.0771$\\
 \hline
 $\delta \angle \text{J}_{4}$  & $0.0005$ & $0.0007$\\ 
 \hline
\end{tabular}
\end{table} 
To understand the role of the ground plane on the modal parameters, two error parameters are defined for the eigenvalues and eigenvectors, respectively. The error parameter for the eigenvalues of the $n^{th}$ mode can be written as
\begin{align}
\delta \lambda_{n}=100|\frac{\lambda_{G,n}-\lambda_{ref.,n}}{\lambda^{c}_{ref.,n}}|\;\%.
\label{sec4_eq36}
\end{align}
The similarity between the reference $n^{th}$ mode and the $n^{th}$ mode with the ground plane is measured by the following error parameter
\begin{align}
\delta \angle \text{I
}_{n}=\cos^{-1}\left(\frac{[\text{I}^c_{G,n}]^{\text{T}}[\text{I}^c_{ref.,n}]}{||[\text{I}^c_{G,n}]||_2||[\text{I}^c_{ref.,n}]||_2}\right) 
\label{sec4_eq37}
\end{align} in degree. The isolated mode parameters $(\lambda_{on},[\text{I}^c_{on}])$ are considered as the reference pair $(\lambda_{ref.,n},[\text{I}^c_{ref.,n}])$.
The characteristic mode variables $(\lambda_{PG,n},[\text{I}^c_{PG,n}])$ can be either with the PEC ground plane $(\lambda_{G,n},[\text{I}^c_{G,n}])$ or with the lossy ground $(\lambda_{LG,n},[\text{I}^c_{LG,n}])$ in \eqref{sec4_eq36} and \eqref{sec4_eq37}. The error parameter $\delta \angle \text{I}_{n}=0$ indicates that the isolated mode and the mode with the ground plane are exactly the same. The error values of the first four dominant modes are compared in Table \ref{Table2} and Table \ref{Table3}. The PEC ground has minimal impact on the modes apart from minor variation in the eigenvalues of the dominant modes. However, the lossy ground has perturbed the modal characteristics for both the eigenvectors and eigenvalues. The perturbation effect seems to be higher for the lower order dominant modes, which are mostly responsible for the radiation from the antenna \cite{sandip_temc1}.

\begin{figure}[ht!]%
\centering
\subfigure[]{%
\label{figIr}
\centering
\includegraphics[height=2.5in, width=2.6in]{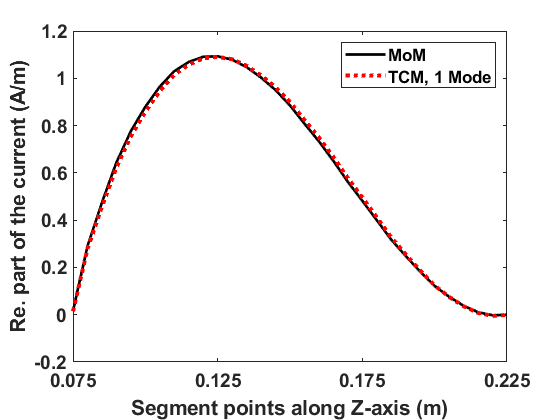}}
\subfigure[]{%
\label{figIx}
\centering
\includegraphics[height=2.5in, width=2.6in]{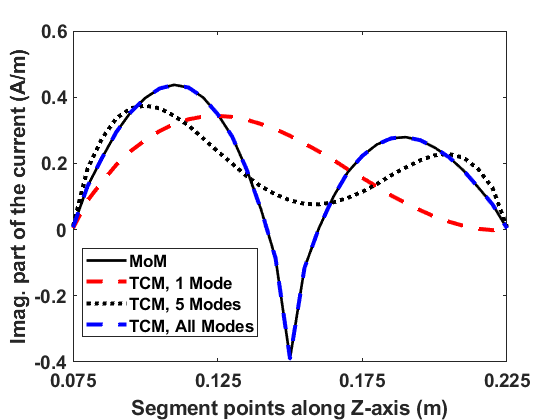}}
\caption{Current comparison for the PEC dipole above the infinite lossy dielectric ground plane.  (a) $\text{J}_{real}$, (b) $\text{J}_{imag}$.}
\end{figure}

To verify the accuracy of the proposed modal approach, the real and imaginary parts of the equivalent current density are compared with the conventional MoM based solutions in Figs. \ref{figIr} and \ref{figIx}. It can be seen from Figs. \ref{figIr} and \ref{figIx} that the real current converges with just the first dominant mode only. However, a slower convergence at the excitation location is noted in the imaginary current. A similar issue was noted for the lossless examples in \cite{cabedo2007theory} and \cite{cabedo2008systematic}. This topic was previously discussed in detail in \cite{TCCM1} and \cite{sghosalthesis}. This topic has been further explored in detail in \cite{sghosalthesis} and \cite{TCCM1}. The field radiated by the structure can be expressed as a weighted sum of the characteristic field patterns \cite{adams2013broadband}.

\begin{figure}[ht!]%
\centering
\subfigure[]{%
\label{figdist}
\centering
\includegraphics[height=2.5in, width=2.6in]{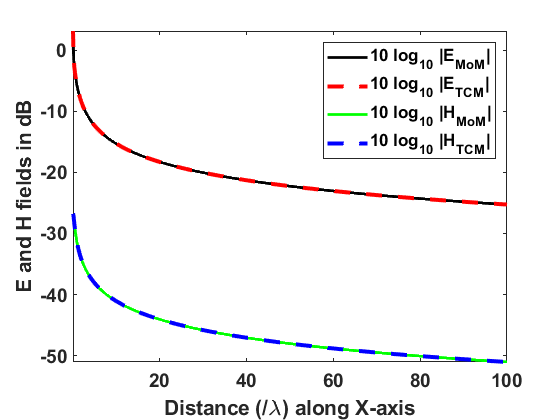}}
\subfigure[]{%
\label{figEV}
\centering
\includegraphics[height=2.5in, width=2.6in]{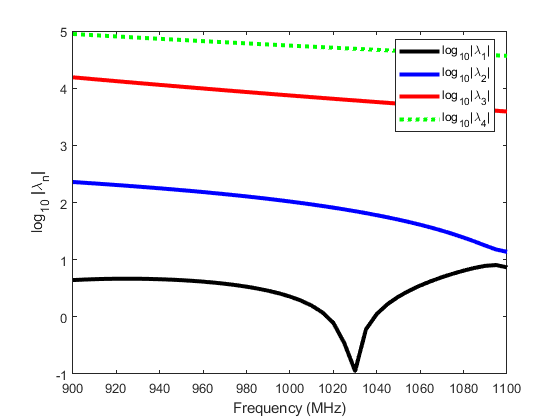}}
\caption{ (a) Transmitted field and (b) eigenvalue comparison for the PEC dipole above the infinite lossy dielectric ground plane.}
\end{figure}

Considering first 5 dominant modes, the electric and magnetic fields have been computed along the X-axis at $0.2\lambda$ height from the ground plane with five dominant modes, and the modal results are found to match closely with the MoM results in Fig. \ref{figdist}. The frequency variation of the eigenvalues of the first five dominant modes can be noted in Fig. \ref{figEV}.
It will be shown in the succeeding discussion on how the vertical gap between the dipole and the dielectric plane controls the modal distribution with respect to the isolated case. The characteristic modes of the isolated element without the ground plane gradually conform to the modes of the element lying above the lossy ground through the perturbation operator of the EFIE equation. 
\color{black}

\newpage

\section{Modal Theory of Ground Coupling Impacts}
Similar to a recently proposed theory of coupled characteristic modes (TCCM) \cite{sghosalthesis} and \cite{TCCM1}, a perturbation approach is introduced here. In the TCCM \cite{TCCM1}, coupling among different antenna elements is analysed. Here, the coupling effects due to the presence of the lossy ground plane have been discussed. It can be seen from \eqref{eqn6} and \eqref{eqn25} that the dimension of the eigenspaces of the isolated dipole and the dipole above the lossy ground are same, i.e., $N \times N$. Therefore, the eigenmode $[\text{J}_{LG,m}]$ has been approximated as a weighted linear combination of the isolated modes $[\text{J}_{on}]$ as
\begin{align}
\text{J}_{LG,m}\cong\sum_{n=1}^N k_{mn}~\text{J}_{on}.
\label{nq1}
\end{align}

The rationale behind such approximation is that only two separate terms are added in \eqref{eqn16} compared to \eqref{eqn2}. This causes a linear translation of the Kernel of the EFIE of the isolated dipole in the free-space. As a consequence, the impedance matrix $[\text{Z}_{LG}]$ is a linear perturbation of the isolated matrix $[\text{Z}_{o}]$ while having the same dimension of eigenspace. On a different note, the alteration of the isolated modes due to the lossy ground plane is a physical phenomenon where the EFIE operator experiences a linear perturbation. Using \eqref{nq1} and \eqref{eqn25}, it can be written that
\begin{align}
\sum_{n=1}^N k_{mn}[\text{X}_{o}][\text{J}_{on}]+\sum_{n=1}^N k_{mn}[\text{X}_{L}][\text{J}_{on}]=\nonumber\\ \lambda_{LG,m}\sum_{n=1}^N k_{mn}[\text{R}_{o}][\text{J}_{on}],
  \label{nq2}
\end{align} 
where $[\text{X}_{L}]=[\text{X}_{LG}]-[\text{X}_{o}]$.
Applying inner product of \eqref{nq1} with the $i^{th}$ isolated mode $[\text{J}_{oi}]$, it will be
\begin{align}
[\text{J}_{oi}]^T\sum_{n=1}^N k_{mn}[\text{X}_{o}][\text{J}_{on}]+[\text{J}_{oi}]^T\sum_{n=1}^N k_{mn}[\text{X}_{L}][\text{J}_{on}]=\nonumber\\ \lambda_{LG,m}[\text{J}_{oi}]^T\sum_{n=1}^N k_{mn}[\text{R}_{o}][\text{J}_{on}].
  \label{nq3}
\end{align} 
The orthogonal properties of the isolated modes of \eqref{eqn6} in \eqref{nq3} will provide
\begin{align}
k_{mi}\lambda_{oi}\text{P}_{Ro,i}+\sum_{n=1}^N k_{mn}\text{P}_{XL,in}=
k_{mi}\lambda_{LG,m}\text{P}_{Ro,i}.
\label{nq4}
\end{align} 
where $\text{P}_{Ro,i}=[\text{J}_{oi}]^T[\text{R}_{o}][\text{J}_{oi}]$ represents the radiative modal power of the $i^{th}$ isolated mode $[\text{J}_{oi}]$. Similarly, the reactive modal power will be $\text{P}_{Xo,i}=[\text{J}_{oi}]^T[\text{X}_{o}][\text{J}_{oi}]=\lambda_{oi}\text{P}_{Ro,i}$. 
The other component $\text{P}_{XL,in}=[\text{J}_{oi}]^T[\text{X}_{L}][\text{J}_{on}]$ denotes the reactive interaction power due to the lossy ground coupling. It can be noted from \eqref{nq3} that the absence of the ground plane indicates that $\lambda_{oi}=\lambda_{LG,m}$. For $i \in [1,N]$, there will be $N$ number of linear equations that will satisfy \eqref{nq4}.The set of all such $N$ linear equations can be expressed as the following matrix form
\begin{align}
\begin{bmatrix}
(\lambda_{LG,m}-\lambda_{o1})\text{P}_{Ro,1} & 0 & \hdots\\
0 & (\lambda_{LG,m}-\lambda_{o2})\text{P}_{Ro,2} & \hdots\\
\vdots & \vdots & \hdots \\
\end{bmatrix}\begin{bmatrix}
k_{m1}\\
k_{m2}\\
\vdots
\end{bmatrix}\nonumber\\=\begin{bmatrix}
\text{P}_{XL,11} & \text{P}_{XL,12} & \hdots\\
\text{P}_{XL,21} & \text{P}_{XL,22} & \hdots\\
\vdots & \vdots & \hdots \\
\end{bmatrix}\begin{bmatrix}
k_{m1}\\
k_{m2}\\
\vdots
\end{bmatrix}
\label{nq5}
\end{align}
As shown in \eqref{nq5}, the left matrix will be a diagonal matrix whereas the right matrix is a dense matrix. As both the matrices $[\text{X}_{LG}]$ and $[\text{X}_{o}]$ are symmetric with the Galerkin's matching, the matrix $[\text{X}_{L}]$ will also be symmetric. Therefore, it can be written that $\text{P}_{XL,in}=[\text{J}_{oi}]^T[\text{X}_{L}][\text{J}_{on}]=[\text{J}_{on}]^T[\text{X}_{L}][\text{J}_{oi}]=\text{P}_{XL,ni}$. With suitable rearrangement, \eqref{nq5} can be written as
\begin{align}
\begin{bmatrix}
(\text{P}_{XL,11}+\text{P}_{Xo,1}) & \text{P}_{XL,12} & \hdots\\
\text{P}_{XL,12} & (\text{P}_{XL,22}+\text{P}_{Xo,2}) & \hdots\\
\vdots & \vdots & \hdots \\
\end{bmatrix}\begin{bmatrix}
k_{m1}\\
k_{m2}\\
\vdots
\end{bmatrix}\nonumber\\=\lambda_{LG,m}\begin{bmatrix}
\text{P}_{Ro,1} & 0 & \hdots\\
0 & \text{P}_{Ro,2} & \hdots\\
\vdots & \vdots & \hdots \\
\end{bmatrix}\begin{bmatrix}
k_{m1}\\
k_{m2}\\
\vdots
\end{bmatrix}
\label{nq6}
\end{align}
It is to be noted here that both the left and right side matrices are real symmetric in \eqref{nq6}. Therefore, it will produce a set of $N$ real eigenvalues $\lambda_{LG,m}$ and $N$ number of eigenvectors, each of which will give the ground coupled mode $\text{J}_{LG,m}$ following \eqref{nq1}. The main contribution of \eqref{nq6} is that it gives a detailed understanding of how the isolated modes are gradually transformed into the ground coupled mode. Similar to \cite{TCCM1}, the general formulation of \eqref{nq6} can be simplified with the closed-form solution for the first order approximation as shown in the next.\\
For the first order approximation, let $\text{J}_{LG,m}$ be approximated as
\begin{align}
\text{J}_{LG,m}=k_1\text{J}_{on_1}+k_2\text{J}_{on_2}.
\label{nq7}
\end{align}
Replacing \eqref{nq7} in \eqref{eqn25}, it will be
\begin{align}
[\text{X}_{LG}](k_1\text{J}_{on_1}+k_2\text{J}_{on_2})&=\lambda_{LG,m}[\text{R}_{o}](k_1\text{J}_{on_1}+k_2\text{J}_{on_2}).
  \label{nq8}
\end{align}
Taking inner product of \eqref{nq8} with $\text{J}_{on_1}$, it will be
\begin{align}
k_1\text{P}_{Xo,n_1}+k_1\text{P}_{XL,n_1 n_1}+k_2\text{P}_{XL,n_1 n_2}=\lambda_{LG,m}k_1\text{P}_{Xo,n_1}.
\label{nq9}
\end{align}
The inner product of \eqref{nq8} with $\text{J}_{on_2}$ provides
\begin{align}
k_1\text{P}_{Xo,n_1}+k_1\text{P}_{XL,n_1 n_1}+k_2\text{P}_{XL,n_1 n_2}=\lambda_{LG,m}k_1\text{P}_{Xo,n_1}.
\label{nq10}
\end{align}
Rearrangement of \eqref{nq9} and \eqref{nq10} will lead to the following quadratic relation
\begin{align}
ak_n^2+bk_n+c=0.
\label{nq11}
\end{align}
where
\begin{align}
k_n=k_2/k_1,~a=\text{P}_{Ro,n_2},~c=\text{P}_{Ro,n_1},\nonumber\\b=
\frac{a(\text{P}_{Xo,n_1}+\text{P}_{XL,n_1n_2})-c(\text{P}_{Xo,n_2}+\text{P}_{XL,n_2n_2})}{\text{P}_{XL,n_1n_2}}.
\label{nq12}
\end{align}
Solution of \eqref{nq11} will be
\begin{align}
k_n=\frac{-b\pm\sqrt{b^2-4ac}}{2a}.
\label{nq13}
\end{align}
Two values of $k_n$ will give two pairs of  $[k_1~~k_2]$, which determine two modes of the dipole above the lossy ground plane. It can be seen in \eqref{nq12} that the parameters $a$ and $c$ depend only on the isolated dipole without the ground plane. The impact of the lossy ground is reflected in $b$ throw the terms $\text{P}_{XL,n_1n_1}$, $\text{P}_{XL,n_2n_2}$, and $\text{P}_{XL,n_1n_2}$. The self-interactions due to the ground plane are indicated by $\text{P}_{XL,n_1n_1}$ and $\text{P}_{XL,n_2n_2}$. Alternately, the mutual interaction of two modes of the isolated dipole due to the presence of the ground plane is represented by $\text{P}_{XL,n_1n_2}$. In the next section, the modes of the same dipole with the same excitation vector will be studied for different separation distances from the lossy dielectric half-space.

\section{Results and Discussions}
For illustration, the same half-wavelength PEC dipole is considered with three different heights from the lossy ground plane as $h=0.3\lambda$, $h=\lambda$, and $h=10\lambda$. Considering the first order approximation of \eqref{nq6} with $n_1=1$ and $n_2=2$, the first two characteristic current modes are computed using the first order approximation as well as using \eqref{eqn25}.\\ 

\begin{table}[ht!]
\centering
\small
\caption{The eigenvalues for the first two dominant modes using the first order approximation.}
\label{Table4}
\begin{tabular}{|c|c|c|c|}
 \hline
 $\lambda_{LG,n}$ & $h=0.3\lambda$  & $h=\lambda$ & $h=10\lambda$\\
 \hline
 $\lambda_{LG,1}$ & $-2.5395$ & $0.6224$ & $0.6256$\\
 \hline
 $\lambda_{LG,2}$ & $-87.6075$ & $-117.9851$ & $-116.8891$\\
 \hline
\end{tabular}
\end{table}
\vspace{-1em}
\begin{table}[ht!]
\centering
\small
\caption{Error of the eigenvalues for the first two dominant modes using the first order approximation.}
\label{Table5}
\begin{tabular}{|c|c|c|c|}
 \hline
 $\delta \lambda_{n}$ & $h=0.3\lambda$  & $h=\lambda$ & $h=10\lambda$\\
 \hline
 $\delta \lambda_{1}$ & $2.0597$ & $0.1752$ & $5.6628E-4$\\
 \hline
 $\delta\lambda_{2}$ & $0.1123$ & $0.0917$ & $9.4384E-5$\\
 \hline
\end{tabular}
\end{table}
\vspace{-1em}
\begin{table}[ht!]
\centering
\small
\caption{Error of the eigenvectors for the first two dominant modes using the first order approximation.}
\label{Table6}
\begin{tabular}{|c|c|c|c|}
 \hline
 $\delta \angle \text{J}_{n}$ & $h=0.3\lambda$  & $h=\lambda$ & $h=10\lambda$\\
 \hline
 $\delta \angle \text{J}_{1}$ & $18.4566$ & $9.0409$ & $0.2697$\\
 \hline
 $\delta \angle \text{I}_{2}$ & $1.0401$ & $1.1191$ & $0.0356$\\
 \hline
\end{tabular}
\end{table} 
\vspace{0.5em}

Corresponding eigenvalues are shown in Table \ref{Table4}. Assuming the modes of \eqref{eqn25} as the reference in \eqref{sec4_eq36} and \eqref{sec4_eq37}, the results with the first order approximation are compared Tables \ref{Table4}--\ref{Table6}. It can be seen from Tables \ref{Table4}--\ref{Table6} that the error amount is relatively high when the dipole is closer to the lossy ground.\\ 
 
\begin{table}[ht!]
\begin{center}
\caption{Comparison of the $\text{P}_{XL,n_1n_2}$ for the three heights}
\label{Table7}
\begin{tabular}{|c|c|c|c|}
 \hline
($n_1,n_2$) & $h=0.3\lambda$  & $h=\lambda$ & $h=10\lambda$\\
 \hline
 (1,1) & $-7.7965E-4$ &$-7.7555E-4$ &$-6.7751E-6$\\
 \hline
 (2,2) & $7.3657E-4$ &$-1.1457E-4$ &$5.2430E-6$\\
 \hline
 (1,2) & $6.5932E-4$ &$-4.4621E-4$ &$-1.3094E-5$\\
 \hline
 \end{tabular}
\end{center}
\end{table}
\vspace{-1em}
 \begin{table}[ht!]
\begin{center}
\caption{Coefficients of the first order approximation for the three heights}
\label{Table8}
\begin{tabular}{|c|c|c|c|}
 \hline
($k_2/k_1$) & $h=0.3\lambda$  & $h=\lambda$ & $h=10\lambda$\\
 \hline
 Mode 1 & $1/0.3281$ &$1/(-0.1575)$ &$1/(-0.0046)$\\
 \hline
 Mode 2 & $(-0.0447)/1$ &$0.0215/1$ &$0.0006/1$\\
  \hline
 \end{tabular}
\end{center}
\end{table}
 
As shown in Tables \ref{Table4}--\ref{Table6}, the impact of the lossy ground tends to decrease as the separation with the ground plane increases. This is due to the lowering of the distance parameters in the Green's function of \eqref{eqn16}. The reason can be more clearly comprehended from Table \ref{Table7} where the self ($\text{P}_{Xo,n_1n_2}, n_1=n_2$) and mutual interaction powers ($\text{P}_{Xo,n_1n_2}, n_1\neq n_2$) 
due to the presence of the lossy ground vary with the separation from the ground plane. As a consequence, the coupling ratio $k_{n2}/k_{n_1}$ or $k_{n1}/k_{n_2}$ of the isolated modes decreases with the height in Table \ref{Table8}. For example, in the case of Mode 1 in Table \ref{Table8}, the contribution from the $2^{nd}$ isolated mode decreases as the separation from the ground increases. Similar nature is observed for Mode 2 too where the weighting coefficient from the $1^{st}$ mode turns out to be only $0.0046$ for $h=10\lambda$.\\
Following the low accuracy of the first order approximation modelling (specifically for the eigenvectors in table \ref{Table6}), the next stage extends the formulation of \eqref{nq6} for the higher order modelling with $4 \times 4$ dimensional interaction matrices as 
\begin{align}
\begin{bmatrix}
\text{P}_{XL,1} & \text{P}_{XL,12}  & \text{P}_{XL,13} & \text{P}_{XL,14}\\
\text{P}_{XL,12} & \text{P}_{XL,2} & \text{P}_{XL,32} & \text{P}_{XL,42}\\
\text{P}_{XL,13} & \text{P}_{XL,23} & \text{P}_{XL,3} & \text{P}_{XL,43} \\
\text{P}_{XL,14} & \text{P}_{XL,24} & \text{P}_{XL,34} &  \text{P}_{XL,4}\\
\end{bmatrix}\begin{bmatrix}
k_{m1}\\
k_{m2}\\
k_{m3}\\
k_{m4}\\
\vdots
\end{bmatrix}\nonumber\\=\lambda_{LG,m}\begin{bmatrix}
\text{P}_{Ro,1} & 0 & 0 & 0\\
0 & \text{P}_{Ro,2} & 0 & 0\\
0 & 0 & \text{P}_{Ro,3} & 0\\
0 & 0 & 0 & \text{P}_{Ro,4}\\
\end{bmatrix}\begin{bmatrix}
k_{m1}\\
k_{m2}\\
k_{m3}\\
k_{m4}\\
\end{bmatrix}
\label{nq14}
\end{align}
where $\text{P}_{XL,n}=\text{P}_{XL,nn}+\text{P}_{Xo,n}$.
\begin{figure}[ht!]%
\centering
\subfigure[]{%
\label{fig2e}%
\centering
\includegraphics[height=2.0in, width=2.2in]{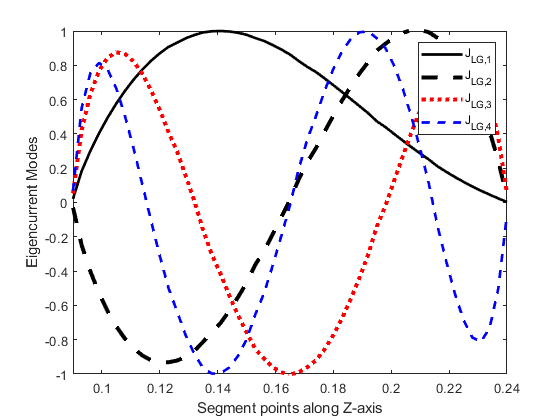}}
\subfigure[]{%
\label{fig2f}%
\centering
\includegraphics[height=2.0in, width=2.2in]{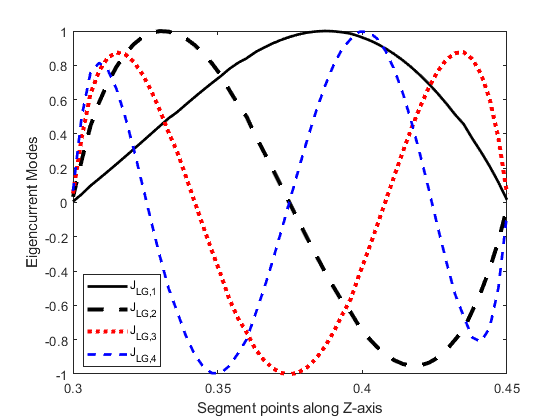}}
\subfigure[]{%
\label{fig2g}%
\centering
\includegraphics[height=2.0in, width=2.2in]{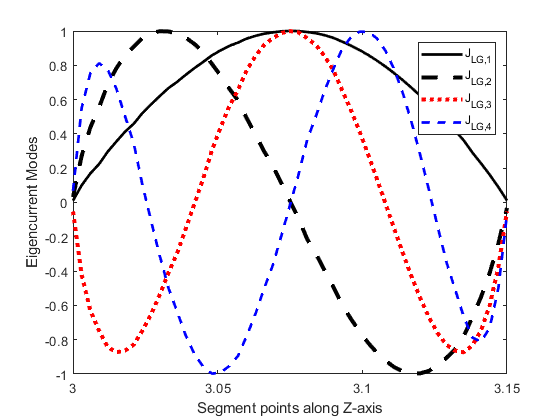}}
\caption{The dipole is lying h meter above the lossy ground plane with $\epsilon_r=16-j16$ and  $L=0.5\lambda$. Normalized distribution of the first four dominant eigencurrent modes considering $4 \times 4$ approximation modelling. (a) $h=0.3\lambda$, (b) $h=\lambda$, (c) $h=10\lambda$.}
\end{figure}

The $4 \times 4$ dimensional generalised characteristic equation of \eqref{nq14} will generate four $4 \times 1$ dimensional eigenvectors that will give four characteristic modes of the dipole above the lossy half-space. It is to be mentioned here that the present manuscript considers the normalisation of the eigenvectors as per the current amplitude based notation of \cite{TCCM1}. In other words, the maximum amplitude of each eigenvector will be $\pm1$. Therefore, the characteristic modes should be normalised compared to its maximum value. Assuming $4 \times 4$ approximation modelling, the first four eigencurrent modes are shown in Figs. \ref{fig2e}--\ref{fig2g}. Corresponding eigenvalues of the first four dominant modes are shown in Table \ref{Table9}. With respect to the reference solution of \eqref{eqn25}, the error parameters are compared in  Tables \ref{Table10} and \ref{Table11}. It can be seen from Tables \ref{Table5} and \ref{Table10} that the predicted eigenvalues with higher order approximation matches closer compared to the first order approximation approach.
\begin{table}[ht!]
\centering
\small
\caption{The eigenvalues for the first four dominant modes using the $4 \times 4$ approximation modelling.}
\label{Table9}
\begin{tabular}{|c|c|c|c|}
 \hline
 $\lambda_{LG,n}$ & $h=0.3\lambda$  & $h=\lambda$ & $h=10\lambda$\\
 \hline
 $\lambda_{LG,1}$ & $-2.4882$ & $0.6235$ & $0.6256$\\
 \hline
 $\lambda_{LG,2}$ & $-87.5092$ & $-117.8770$ & $-116.8890$\\
 \hline
 $\lambda_{LG,3}$ & $-7.4773E+3$ & $-7.3711E+3$ & $-7.3843E+3$\\
 \hline
 $\lambda_{LG,4}$ & $-7.9487E+5$ & $-7.9510E+5$ & $-7.9504E+5$\\
 \hline
\end{tabular}
\end{table}
\vspace{-1em}
\begin{table}[ht!]
\centering
\small
\caption{Error of the eigenvalues for the three cases with the $4 \times 4$ approximation modelling.}
\label{Table10}
\begin{tabular}{|c|c|c|c|}
 \hline
 $\delta \lambda_{n}$ & $h=0.3\lambda$  & $h=\lambda$ & $h=10\lambda$\\
 \hline
 $\delta \lambda_{1}$ & $ 7.0874E-4$ & $2.3857E-4$ & $ 1.4352E-7$\\
 \hline
 $\delta\lambda_{2}$ & $9.4707E-6$ & $1.5926E-5$ & $2.6148E-8$\\
 \hline
  $\delta\lambda_{3}$  & $2.7576E-6$ & $3.5791E-7$ & $2.0359E-10$\\
 \hline
 $\delta\lambda_{4}$  & $7.5724E-9$ & $7.9950E-9$ & $3.5347E-11$\\
 \hline
\end{tabular}
\end{table}
\vspace{-1em}
\begin{table}[ht!]
\centering
\small
\caption{Error of the eigenvectors for the three cases with the $4 \times 4$ approximation modelling.}
\label{Table11}
\begin{tabular}{|c|c|c|c|}
 \hline
 $\delta \angle \text{J}_{n}$ & $h=0.3\lambda$  & $h=\lambda$ & $h=10\lambda$\\
 \hline
 $\delta \angle \text{J}_{1}$ & $0.0229$ & $0.0064$ & $1.4139E-4$\\
 \hline
 $\delta \angle \text{I}_{2}$ & $0.0018$ & $0.0090$ & $3.6739E-4$\\
 \hline
  $\delta \angle\text{J}_{3}$  & $0.0052$ &  $0.0018$ & $3.6673E-5$\\
 \hline
 $\delta \angle \text{J}_{4}$  & $3.6627E-4$ & $4.1416E-4$ & $2.3845E-5$\\ 
 \hline
\end{tabular}
\end{table} 
\begin{table}[ht!]
\centering
\small
\caption{Error of the eigenvalues for the three cases with the full $N \times N$ approximation modelling.}
\label{Table12}
\begin{tabular}{|c|c|c|c|}
 \hline
 $\delta \lambda_{n}$ & $h=0.3\lambda$  & $h=\lambda$ & $h=10\lambda$\\
 \hline
 $\delta \lambda_{1}$ & $2.1635E-10$ & $8.8828E-10$ & $1.8520E-9$\\
 \hline
 $\delta\lambda_{2}$ & $9.4513E-12$ & $2.5558E-12$ & $9.9692E-13$\\
 \hline
  $\delta\lambda_{3}$  & $1.6785E-12$ & $2.6898E-12$ & $6.8973E-13$\\
 \hline
 $\delta\lambda_{4}$  & $2.7388E{-12}$ & $4.1289E-12$ & $8.9320E-13$\\
 \hline
\end{tabular}
\end{table}
\vspace{-1em}
\begin{table}[ht!]
\centering
\small
\caption{Error of the eigenvectors for the three cases with the full $N \times N$ approximation modelling.}
\label{Table13}
\begin{tabular}{|c|c|c|c|}
 \hline
 $\delta \angle \text{J}_{n}$ & $h=0.3\lambda$  & $h=\lambda$ & $h=10\lambda$\\
 \hline
 $\delta \angle \text{J}_{1}$ & $0$ & $1.2074E-6$ & $1.2074E-6$\\
 \hline
 $\delta \angle \text{I}_{2}$ & $0$ & $1.2076E-6$ & $0$\\
 \hline
  $\delta \angle\text{J}_{3}$  & $0$ &  $0$ & $1.2074E-6$\\
 \hline
 $\delta \angle \text{J}_{4}$  & $0$ & $0$ & $0$\\ 
 \hline
\end{tabular}
\end{table} 
\vspace{0.5em}

The accuracy of the predicted eigenvectors improves significantly with the higher order approximation as shown in Tables \ref{Table6} and \ref{Table11}. If the full $N \times N$ matrix solution of \eqref{nq6} is considered, the eigenvalues and eigenvectors become closely the same as of \eqref{eqn25} as shown in Tables \ref{Table12} and \ref{Table13}. This justifies the theory of linear combination of \eqref{nq1}.\\
\begin{align}
\eta_{LG}=\frac{\text{P}_{R,LG}}{\text{P}_{R,iso}}=\frac{Re.([\text{J}_{LG}]^H[\text{Z}_{LG}][\text{J}_{LG}])}{Re.([\text{J}_{o}]^H[\text{Z}_{o}][\text{J}_{o}])}=\frac{\text{R}^{rad}_{LG}}{\text{R}^{rad}_{o}}.
\label{eg1}
\end{align}
where \enquote{$H$} stands for the complex conjugation operation. The numerator represents the radiated power of the isolated dipole without the ground plane, and the denominator indicates the radiated power of the dipole above the lossy ground plane. The radiation resistance for the isolated and with the lossy ground cases are denoted by $\text{R}^{rad}_{LG}$ and $\text{R}^{rad}_{o}$, respectively. For the isolated dipole, total radiated power can be written in terms of the modal parameters as \cite{sandip_temc1}
\begin{align}
\text{P}_{R,iso}=\sum_{n=1}^N |\alpha_{on}|^2 [\text{J}_{on}]^T[\text{R}_{o}][\text{J}_{on}],
\label{eg2}
\end{align}
where
\begin{align}
\alpha_{o,n}=&=\frac{[\text{J}_{o,n}]^T[\text{V}]}{(1+\text{j}\lambda_{o,n})[\text{J}_{o,n}]^T[\text{R}_{o}][\text{J}_{o,n}]}.
\label{eg4}
\end{align} 
Similarly, the radiated power for the dipole above the lossy ground plane will be
\begin{align}
\text{P}_{R,LG}=\sum_{n=1}^N |\alpha_{LG,n}|^2 [\text{J}_{LG,n}]^T[\text{R}_{o}][\text{J}_{LG,n}].
\label{eg3}
\end{align} 

\begin{table}[ht!]
\begin{center}
\caption{Comparison of $\eta_g$ for the three heights}
\label{Table14}
\begin{tabular}{|c|c|c|c|}
 \hline
$h=0.3\lambda$  & $h=\lambda$ & $h=10\lambda$ & $h=300\lambda$\\
 \hline
$0.1337$ &$0.7134$ &$0.9148$ & $0.9998$\\
 \hline
 \end{tabular}
\end{center}
\end{table}

The efficiency parameters for different heights from the lossy ground are shown in Table \ref{Table14}. It can be seen from Table \ref{Table14} that the dipole with larger separation possesses $\eta_{LG}$ closer to 1. This indicates that the modes approach towards the isolated case. As the modal current distribution varies with ground separation, corresponding transmitted field patterns will also vary with the vertical gap. From a practical perspective, it can be useful in the antenna positioning for transmitting the desired pattern in the presence of the lossy ground plane \cite{sghosalthesis}.

\newpage
\section{Conclusion}
\color{black}Following the limitation of the conventional characteristic mode analysis, the present work proposed a new eigenmode formulation for the vertical dipole lying above the lossy dielectric half-space. Three different case studies were considered for comparison-an isolated dipole, a vertical dipole above the infinite PEC ground plane, and a vertical dipole above the infinite lossy ground plane. A systematic blending of the theory of complex image and the theory of characteristic modes of PEC antennas provided the motivation of the proposed modal analysis. Later, a linear combination of the isolated modes were used to identify the underlying physics behind the impacts of the lossy ground on the modal distribution. The proposed theory can find various applications like in lightning field analysis, wireless propagation modelling where the radiating sources lie above the lossy ground plane.

\newpage
\appendices

\section{Modal Formulations for the Lossless Cases}
The MoM impedance matrix $[\text{Z}_0]=[\text{R}_0]+\text{j}[\text{X}_0]$ for the PEC dipole lying in the free-space needs to consider the following Green's function  \cite{harrington1993field}
\begin{align}
    G_o(z,z')&=\frac{e^{-jk_0 d_o}}{4\pi d_o} \label{eqn2}
\end{align}
The characteristic modes of the isolated dipole can be computed as \cite{harrington1971theory}
\begin{align}
[\text{X}_{o}]\text{J}_{on}=\lambda_{on}[\text{R}_{o}]\text{J}_{on}.
\label{eqn6}
\end{align} 
Similarly, the Green's function to compute the MoM impedance matrix $[\text{Z}_{PG}]=[\text{R}_{PG}]+\text{j}[\text{X}_{PG}]$ for the PEC dipole lying above the PEC half-space will be 
\begin{align}
    G_{PG}(z,z')&=\frac{e^{-jk_0d_o}}{4\pi R_o}+\frac{e^{-jk_0 d_{i}}}{4\pi d_{i}}
 \label{eqn9}
\end{align} 
The characteristic mode for this case can be determined as \cite{harrington1971theory}
\begin{align}
[\text{X}_{PG}]\text{J}_{PG,n}=\lambda_{PG,n}[\text{R}_{PG}]\text{J}_{PG,n}.
\label{eqn13}
\end{align} 

\color{black} 
\newpage
\section*{Acknowledgement}
The authors would like to thanks the reviewers and editors for their valuable suggestions to improve this manuscript.  Thanks to Dr. Rakesh Sinha for several fruitful discussions on the characteristic mode analysis. The work was carried out in the Department of Electronics and Electrical Communication Engineering, Indian Institute of Technology Kharagpur, Kharagpur, India.

\newpage
\bibliographystyle{IEEEtran}
\bibliography{mainref}

\end{document}